\documentclass[fp]{jpsj3}
\usepackage[]{graphicx,color}

\newcommand{\llangle}{\left\langle\!\!\left\langle}
\newcommand{\rrangle}{\right\rangle\!\!\right\rangle}

\begin{document}
\title{
First-principles study of 
intersite magnetic couplings
and Curie temperature in
RFe$_{12-x}$Cr$_{x}$ (R = Y, Nd, Sm)
}
\author{Taro Fukazawa$^{1,3}$, Hisazumi Akai$^{2,3}$,
Yosuke Harashima$^{1,3}$ and  Takashi Miyake$^{1,3}$}

\inst{$^1$CD-FMat, National Institute of Advanced Industrial Science
and Technology, Tsukuba, Ibaraki 305-8568, Japan \\
$^2$The Institute for Solid State Physics, The University of
Tokyo, 5-1-5 Kashiwano-ha, Kashiwa, Chiba 277-8581, Japan \\
$^3$ESICMM, National Institute for Materials Science,
Tsukuba, Ibaraki 305-0047, Japan}

\date{\today}
\abst{
 We present a first-principles study of
 RFe$_{12-x}$Cr$_{x}$ (R = Y, Nd, Sm)
 crystals with ThMn$_{12}$ structure.
 We discuss, within the mean field approximation,
 intersite magnetic couplings calculated
 using Liechtenstein's formula
 and convert them into Curie temperatures,
 $T_{\rm C}$,
 which are found to become larger
 when a small amount of Cr ($x \leq 0.5$) is introduced
 into the system. This enhancement
 is larger than that for Co in the dilute limit,
 $x \rightarrow 0$.
 In contrast, above $x > 0.5$, the Curie temperature decreases
 as Cr concentration increases.
 This behavior is analyzed using an
 expansion of $T_{\rm C}$ in terms of concentration.
}
\maketitle

\section{Introduction}
Iron compounds with the ThMn$_{12}$ structure
[space group: I4/mmm (\#139)]
are considered to be
a candidate for the main phase of permanent magnets that,
 because of their high Fe content, 
can surpass the Nd$_2$Fe$_{14}$B magnet in quality, especially in magnetization.
Successful synthesis of SmFe$_{12}$ and SmFe$_{12}$N as films was reported
in Wang et al.\cite{Wang93}
Hirayama et al. reported the synthesis of NdFe$_{12}$ and NdFe$_{12}$N films, 
and the high magnetization and anisotropy field 
of NdFe$_{12}$N.\cite{Hirayama15,Hirayama15b}
However, their Curie temperatures
are not as high as 
previously expected.\cite{Buschow91}

Some of the RFe$_{12-x}$M$_x$ compounds (R: rare earth; M: metal)
are thermally much more stable than RFe$_{12}$ 
and hence they can exist as bulk material.
Introducing M for stabilization can affect
their Curie temperature, $T_{\rm C}$,
which has been investigated by several authors.\cite{Buschow91}
However, the range in concentration of M observed in experiments
was limited at the time because of the thermal instability.

Ogura et al.\cite{Ogura11} discussed the addition of Cr and V into iron systems based on a first-principles calculation.
They showed that the Curie temperature is enhanced by Cr in a hypothetical Fe$_{15}$Cr system, which they attributed to Cr around which the local electronic structure of the nearest Fe atoms became Co-like.
They also suggested that Fe/Cr heterostructures could achieve higher $T_{\rm C}$ than the pure Fe system.

In regard to rare-earth permanent magnets, the enhancement of the Curie temperature by doping with Cr has been experimentally observed in 2--17 systems:
Hao et al.\cite{Hao96} showed that Th$_2$Ni$_{17}$-type Y$_{2}$Fe$_{17-x}$Cr$_x$
has a $\sim$100~K higher value of $T_{\rm C}$ at $x=1$ than $x=0$;
Girt et al.\cite{Girt97} showed that Th$_2$Zn$_{17}$-type
Nd$_{2}$Fe$_{17-x}$Cr$_x$ has a $\sim$50~K higher value of $T_{\rm C}$ at $x=1$ than $x=0$.
Both attribute the enhancement to a decrease in the anti-ferromagnetic coupling between the shortest Fe--Fe bonds in the system through the substitution of Cr.

In this study, we investigate RFe$_{12-x}$Cr$_x$ for R= Y, Nd, Sm. 
We discuss intersite magnetic couplings 
calculated following Liechtenstein's method.\cite{Liechtenstein87}
The value of each of these couplings is converted to a Curie temperature
using the mean field approximation.
The calculated $T_{\rm C}$ is enhanced by Cr in the concentration range
$0 < x \leq 0.5$, for which there have been no experimental 
reports of $T_{\rm C}$ to the best of our knowledge.
This enhancement induced by Cr is larger than that by Co in this regime.
However, at a certain concentration in $x > 0.5$, the Curie temperature begins 
to decrease as the Cr concentration increases.
This non-linear behavior of $T_{\rm C}$ as a function of $x$
is analyzed using a concentration expansion of $T_{\rm C}$, 
and explained in terms of inter-sublattice couplings
for Fe--Fe, Fe--Cr, and Cr--Cr.

\section{Methods}
We use the Korringa--Kohn--Rostoker Green function method
for solving the Kohn--Sham equation\cite{Kohn65}
obtained from density functional theory.\cite{Hohenberg64}
The local density approximation is used in the calculation;
the spin--orbit coupling at the R site is taken into account with the f-electrons
treated as a trivalent open core for which the configuration is constrained by Hund's rule;
the self-interaction correction\cite{Perdew81} is also applied to the f-orbitals.
Fe and the dopants are treated within the coherent potential
approximation (CPA) under the assumption that Cr (or Co) occupies 
Fe(8j), Fe(8i), and Fe(8f) sites with an equal probability.
We refer readers to Ref.~\citen{Fukazawa17} for further details of the calculation setup.

We use the lattice parameters of RFe$_{12}$ obtained using QMAS,\cite{QMAS}
which is based on the PAW method,\cite{Bloechl94,Kresse99}
within a generalized gradient approximation for the calculations 
involving the RFe$_{12-x}$M$_{x}$ (M=Cr, Co) system.
We refer readers to Ref.~\citen{Harashima14b} for details of the calculation setup.
These values of the lattice parameters are given in Appendix \ref{lattparams}.

The values of intersite magnetic couplings, $J^\text{A--B}_{i,j}$,
are calculated using Liechtenstein's formula.\cite{Liechtenstein87}
These values are obtained within perturbation theory from energy shifts due to spin rotation of atom A placed at the $i$th site and atom B placed at the $j$th site in the environment of the coherent potentials.

In our calculation of the Curie temperature,
we considered a sample-dependent Heisenberg-like Hamiltonian $\mathcal{H}(n)$
for the $n$th sample ($n=1,2,\cdots$) in the form of
\begin{equation}
 \mathcal{H}(n)
  =
  -
  \sum_{i}
  \sum_{j}
  J_{i, j}(n)\,
  \vec{e}_{i,n}
  \cdot
  \vec{e}_{j,n}
  \label{HeisenbergHamiltonian}
\end{equation}
where $\vec{e}_{i,n}$ denotes a unit vector in the direction of the local spin-polarization
at the $i$th site of the $n$th sample, and $J_{i, j}$ is a random coupling made of
$J^{\text{A--B}}_{i,j}$ determined by the Liechtenstein formula:
\begin{equation}
 J_{i, j}(n)\,
  =
  \sum_{\rm A, B}
  \gamma^{\rm A}_i(n)
  \,
  \gamma^{\rm B}_j(n)
  \,
  J^{\text {A--B}}_{i, j}\,
  \label{RandomizedJij}
\end{equation}
where $\gamma^{\rm A}_i(n)$ is a random variable corresponding to
the occupation number of the A atom at the $i$th site of the $n$th sample
(therefore the value of $\gamma^{\rm A}_i(n)$ must be either 0 or 1).
We assume quenched randomness for the systems, and
the occupation number at a site is considered independent of that at other sites.
The distribution of $\gamma^{\rm A}_i(n)$ is taken so that
its sample average, $C^{\rm A}_i$, becomes
the concentration assumed in the KKR-CPA calculation.
Specifically, for the current case, 
$C^{\rm R}=1$, $C^{\rm Fe}=1-x/12$, and $C^{\rm Cr}=x/12$.
Based on this Hamiltonian, the Curie temperature is estimated
using the mean-field approximation, which is summarized in Appendix \ref{section_MFA}.

\section{Results and Discussion}
\begin{figure}[h]
 \centering
 \includegraphics[width=8cm]{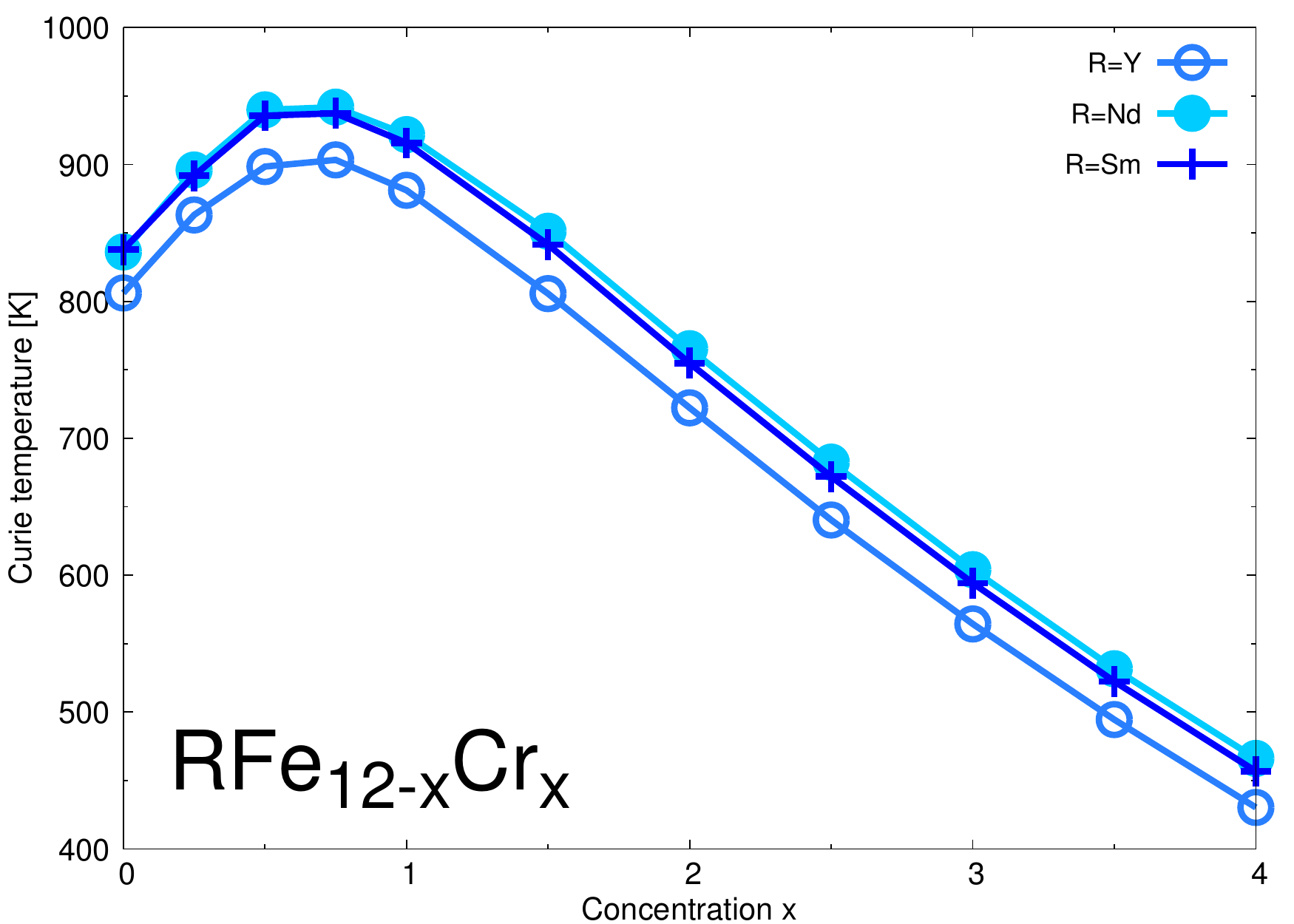}
 \caption{(Color online)
 Values of the Curie temperature for RFe$_{12-x}$Cr$_x$ (R = Y, Nd, Sm)
 as functions of Cr concentration $x$.
 \label{Tc_RFeM12}
}
\end{figure}

\begin{figure}[h]
  \center
  \includegraphics[width=8cm]{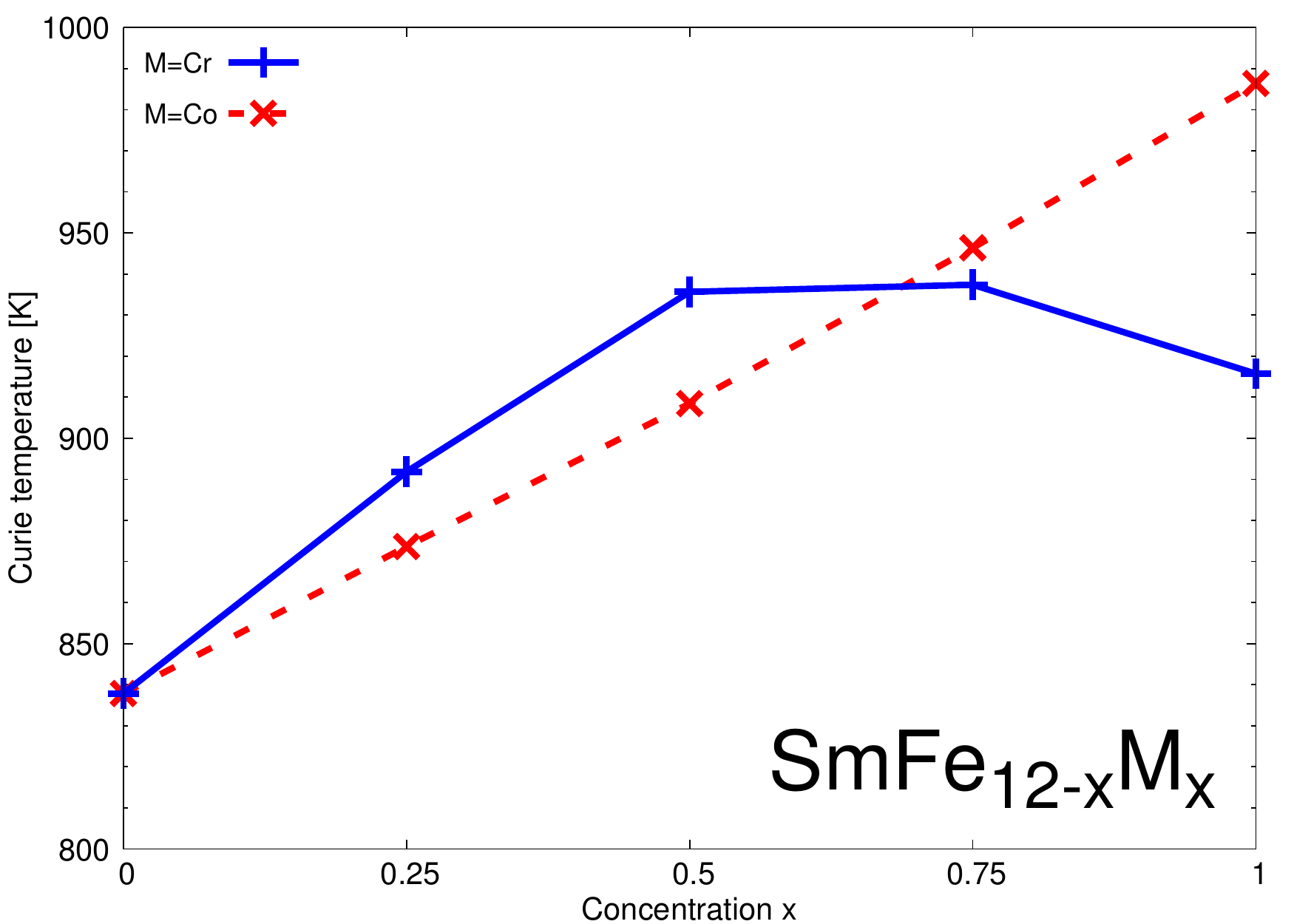}
  \vspace{1cm}
  \caption{(Color online)
  Values of the Curie temperature for SmFe$_{12-x}$Cr$_x$
  compared with those for SmFe$_{12-x}$Co$_x$ in the range $0\leq x \leq 1$
  as functions of Cr concentration $x$.
  \label{Tc_SmFeM12}
  }
 \end{figure}

Figure \ref{Tc_RFeM12} shows the values of the calculated Curie temperature, $T_\text{C}$,
for RFe$_{12-x}$Cr$_x$ as functions of $x$ for $0 \leq x \leq 4$.
The values of $T_\text{C}$ increase as the concentration of Cr increases from $x=0$ in the range $0<x\leq 0.5$.
As the Cr concentration increases from $x=0.75$, $T_\text{C}$ values decrease.
Although our values are much higher than the experimental values
(cf. $T^{\rm exp.}_\text{C}$ = 593 K\cite{Wang93} for SmFe$_{12}$; $T^{\rm exp.}_\text{C}$ = 555 K\cite{Hirayama17} also for SmFe$_{12}$), 
mainly because the mean-field approximation is used,
the linear behavior of the experimental curve\cite{Buschow91}
for $x>1$ is well reproduced.
It is also shown later that the calculated Curie temperature of Sm(Fe$_{12-x}$Co$_x$)
is increased by $\sim 150$ K (Fig.~\ref{Tc_SmFeM12}) from $x=0$ to $x=1$.
This value is comparable to the increase of 155~K from $x=0$ to $x=1.2$
in an experiment.\cite{Hirayama17}
Comparison between calculated and experimental $T_{\rm C}$ within the mean-field approximation
for other ThMn$_{12}$-type systems are also presented in our previous paper,
and they are also in good agreement when a relative change is considered.\cite{Fukazawa17}

Let us compare the enhancement in $T_\text{C}$ caused by Cr with that by Co
because Co is a typical element that is commonly used for increasing
the Curie temperature for Fe-based systems.
From here on, we take SmFe$_{12-x}$Cr$_{x}$ as typical of
the RFe$_{12-x}$Cr$_{x}$ systems; this is justified from 
the strong resemblance of all the $T_\text{C}$ curves. 
Figure \ref{Tc_SmFeM12} compares $T_\text{C}$
for SmFe$_{12-x}$Cr$_{x}$ as a function of $x$
and that for SmFe$_{12-x}$Co$_{x}$ in the range $0\leq x \leq 1$.
The enhancement in $T_\text{C}$ by Cr is stronger than that by Co.

To analyze the behavior of the $T_\text{C}$ curves in the dilute region,
we consider the concentration expansion of $T_\text{C}$:
\begin{equation}
 T_\text{C}(x)
  =
  T_\text{C}(0)
  +
  \alpha_1 x
  +
  \alpha_2 x^2
  +
  \cdots.
  \label{Tc_expansion}
\end{equation}
The difference in rise between M=Cr and M=Co can be attributed to 
the difference in $\alpha_1$, which is the derivative of the $T_\text{C}(x)$ at $x=0$.
Within the mean-field approximation, the Curie temperature depends
on the intersite magnetic couplings $\{J^\text{A--B}_{ij}\}$
and the average values of the occupation numbers, $\{C^\text{A}_{i}\}$.
Because $\{J^\text{A--B}_{ij}\}$ and $\{C^\text{A}_{i}\}$
depend on $x$, $\alpha_1(=dT_\text{C}/dx)$ can be written as 
a sum of partial derivatives with respect to them.
Then, $\alpha_1$ can be written as
\begin{align}
  \alpha_1
  &=
  \left.
  \frac
  {dT_\text{C}
      \left[
          \{C^\text{A}_{i}(x)\},\{J^\text{A--B}_{ij}(x)\}
      \right]
  }
  {dx}
  \right|_{x=0}
  \nonumber \\
  &=
  \sum_{i}
  \sum_\text{A}
  \left.
  \frac{\partial T_\text{C}}{\partial C^\text{A}_{i}}
  \right|_{\{J^\text{A--B}_{ij}(0)\}}
  \left.
  \frac{dC^\text{A}_{i}}{dx}
  \right|_{x=0}   \nonumber \\
  &+
  \sum_{i,j}
  \sum_\text{A,B}
  \left.
  \frac{\partial T_\text{C}}{\partial J^\text{A--B}_{ij}}
  \right|_{\{C^\text{A}_{i}(0)\}}
  \left.
  \frac{dJ^\text{A--B}_{ij}}{dx}
  \right|_{x=0}
 .
 \label{alpha_decomp}
\end{align}

Calling the first and second terms in the final expression as the ``Direct'' and the ``Indirect'' parts,
the former represents the direct effect obtained by replacing the Fe--Fe bonds with Fe--Cr bonds, 
and derives solely from the difference in the couplings associated with the replaced bonds and its substitute, $J^\text{Fe--Cr}_{i,j}-J^\text{Fe--Fe}_{i,j}$
(and $J^\text{Cr--Fe}_{i,j}-J^\text{Fe--Fe}_{i,j}$) for $x=0$.
The Indirect part represents the influence of the replacement on the remaining Fe--Fe couplings, 
and includes Cr's effect in making the surrounding Fe atoms
appear Co-like as discussed by Ogura et al.\cite{Ogura11}.

Table \ref{table:dTc/dx} lists the values of $dT_\text{C}/dx$
for RFe$_{12-x}$Cr$_{x}$ and RFe$_{12-x}$Co$_{x}$, and how 
they are decomposed into Direct and Indirect parts.
We performed the calculation for five concentrations in the range 
$x=0\operatorname{--}0.05$ to obtain their values.
The values of $dT_\text{C}/dx$ for the Cr systems are significantly 
larger than those for the Co systems as expected.
Values of the Direct and Indirect parts have similar magnitude.
Therefore, the replacement of Fe--Fe bonds with Fe--Cr is as important as Cr's effect 
in making surrounding Fe atoms appear Co-like in the enhancement of $T_\text{C}$.
\begin{table}[ht]
\center
\caption{ 
 Values of $dT_\text{C}/dx$ at $x=0$ for RFe$_{12-x}$Cr$_x$ and RFe$_{12-x}$Co$_x$,
 and their decomposition into ``Direct'' and ``Indirect'' parts of Eq.~\eqref{alpha_decomp}.
 \label{table:dTc/dx}}
\begin{tabular}{cccc}
\hline
\hline
 & $dT_\text{C}/dx$ [K] & Direct [K] & Indirect [K] \\
\hline
  YFe$_{12-x}$Cr$_{x}$ & 304 & 158 & 146 \\
 NdFe$_{12-x}$Cr$_{x}$ & 335 & 173 & 162 \\
 SmFe$_{12-x}$Cr$_{x}$ & 321 & 169 & 152 \\
\hline
  YFe$_{12-x}$Co$_{x}$ & 148 &  75 &  73 \\
 NdFe$_{12-x}$Co$_{x}$ & 157 &  87 &  70 \\
 SmFe$_{12-x}$Co$_{x}$ & 157 &  87 &  71 \\
\hline
\hline
\end{tabular}
\end{table}

To provide a quick comparison of the Fe--Cr couplings with other types of couplings, we use the summation of $J^\text{A--B}_{ij}$ defined by
\begin{equation}
 \mathcal{J}_{\text{A}(\mu)\text{--B}(\nu)}
  =
  \sum_{j\in\text{B}(\nu)}
  J^\text{A--B}_{ij}
  \quad
  \left[ i \in \text{A}(\mu)\right],
\end{equation}
where A($\mu$) and B($\nu$) denote the sub-lattices composed of A atoms at the $\mu$ sites and
B atoms at the $\nu$ sites, respectively (A, B = Fe or Cr; $\mu,\nu$ = 8f, 8i, 8j).
In the following, we consider the sub-lattices composed of Fe atoms
and the sub-lattice composed of Cr atoms, separately.
With this set-up, $\mathcal{J}_{\text{A}(\mu)\text{--B}(\nu)}$ holds
$\mathcal{J}_{\text{A}(\mu)\text{--B}(\nu)}
=\mathcal{J}_{\text{B}(\nu)\text{--A}(\mu)}$ when A, B $\in$
\{Fe, Cr\}. Therefore, there are six different
$\mathcal{J}_{\text{Fe}(\mu)\text{--Fe}(\nu)}$'s,
six different $\mathcal{J}_{\text{Cr}(\mu)\text{--Cr}(\nu)}$'s,
and nine different $\mathcal{J}_{\text{Fe}(\mu)\text{--Cr}(\nu)}$'s.
To further simplify the analysis, we average these $\mathcal{J}$'s
into $\mathcal{J}_{\text{Fe}\text{--Fe}}$,
$\mathcal{J}_{\text{Cr}\text{--Cr}}$, and 
$\mathcal{J}_{\text{Fe}\text{--Cr}}$.
The same averaging is also performed for the Co systems.

\begin{figure}[h]
  \center
  \includegraphics[width=8cm]{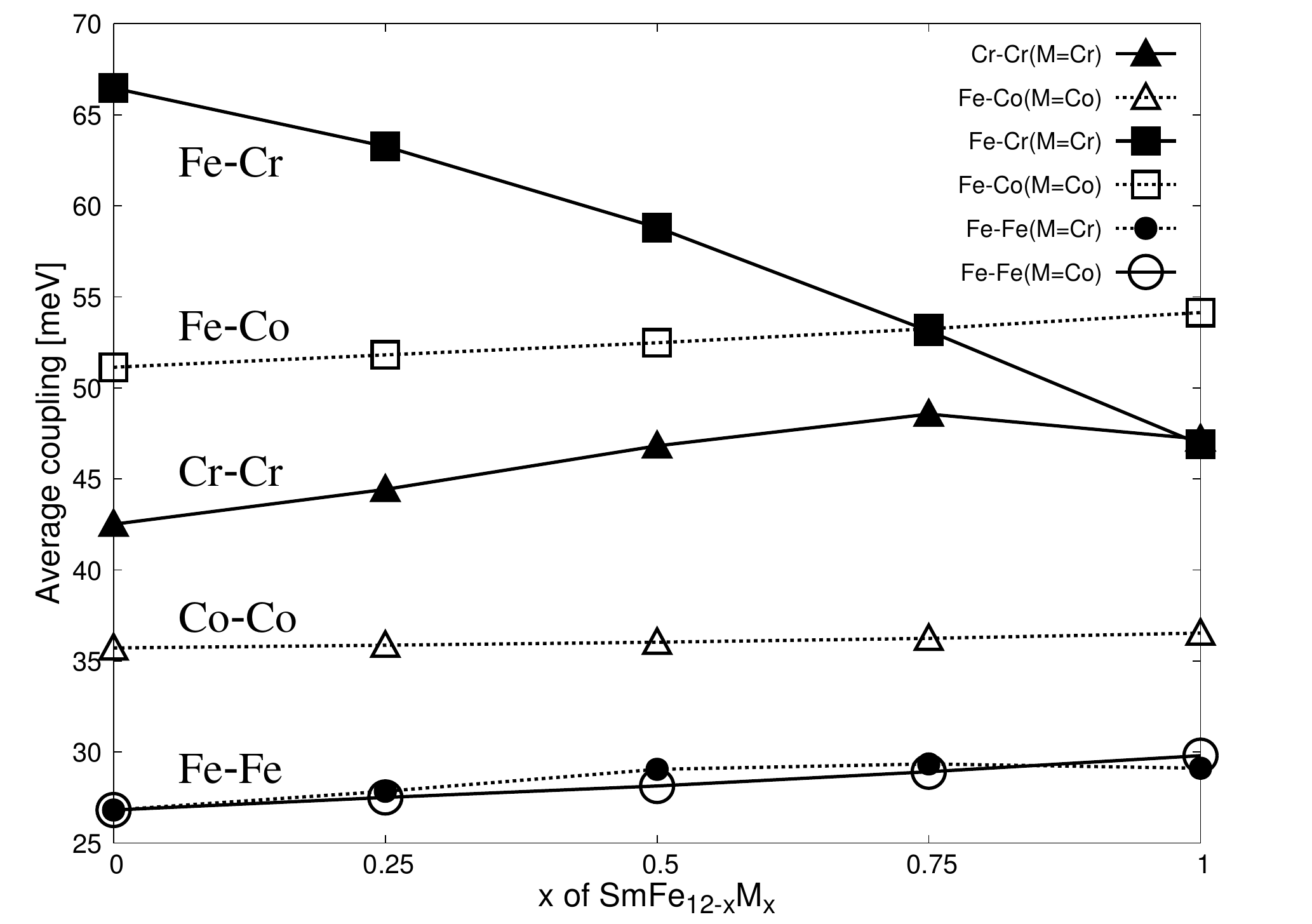}
  \vspace{1cm}
  \caption{
  Absolute values of $\mathcal{J}_{\text{Cr}\text{--Cr}}$,
  $\mathcal{J}_{\text{Fe}\text{--Cr}}$ and 
  $\mathcal{J}_{\text{Fe}\text{--Fe}}$ in SmFe$_{12-x}$Cr$_x$
  as functions of concentration $x$, and those of
  $\mathcal{J}_{\text{Co}\text{--Co}}$,
  $\mathcal{J}_{\text{Fe}\text{--Co}}$ and 
  $\mathcal{J}_{\text{Fe}\text{--Fe}}$ in SmFe$_{12-x}$Co$_x$.
  \label{J_ave}
  }
 \end{figure}

Figure \ref{J_ave} shows absolute values of 
$\mathcal{J}_{\text{Cr}\text{--Cr}}$,
$\mathcal{J}_{\text{Fe}\text{--Cr}}$,
 and 
$\mathcal{J}_{\text{Fe}\text{--Fe}}$ for SmFe$_{12-x}$Cr$_x$, and 
$\mathcal{J}_{\text{Co}\text{--Co}}$,
$\mathcal{J}_{\text{Fe}\text{--Co}}$ and 
$\mathcal{J}_{\text{Fe}\text{--Fe}}$ for SmFe$_{12-x}$Co$_x$
as functions of concentration $x$.
The values of $\mathcal{J}_{\text{Cr}\text{--Cr}}$ and 
$\mathcal{J}_{\text{Fe}\text{--Cr}}$ are negative (antiferromagnetic) and 
those of $\mathcal{J}_{\text{Co}\text{--Co}}$,
$\mathcal{J}_{\text{Fe}\text{--Co}}$ and $\mathcal{J}_{\text{Fe}\text{--Fe}}$
are positive (ferromagnetic).
The absolute value of $\mathcal{J}_{\text{Fe}\text{--Cr}}$ is significantly 
larger than $\mathcal{J}_{\text{Fe}\text{--Co}}$ and 
$\mathcal{J}_{\text{Fe}\text{--Fe}}$ at $x=0$,
which means the antiferromagnetic coupling between Fe and Cr
is stronger than the Fe--Fe and Fe--Co coupling,
and the Fe--Cr coupling stabilizes the ground state in the $x \ll 1$ region.
However, this Fe--Cr coupling becomes weaker as the concentration of Cr increases.
Also it has almost the same value with the Fe--Co couplings
at $x=0.75$, and becomes smaller at $x=1$,
which corresponds well to the crossing of the two curves in Fig.~\ref{Tc_SmFeM12}.
This weakening of the Fe--Cr couplings significantly contributes to 
the quadratic terms in Eq.~\eqref{Tc_expansion}.
Although it does not affect the behavior of $T_{\rm C}$
to first order, this produces a quadratic behavior in $T_{\rm C}$ curve 
very quickly as the concentration $x$ increases.
It is also noteworthy that the value of $\mathcal{J}_{\text{Cr}\text{--Cr}}$ is negative and the Cr--Cr antiferromagnetic couplings also contribute significant quadratic terms that reduce the Curie temperature as these couplings are against the spin-alignment of the ground state.

\begin{figure}[h]
  \center
  \includegraphics[width=8cm]{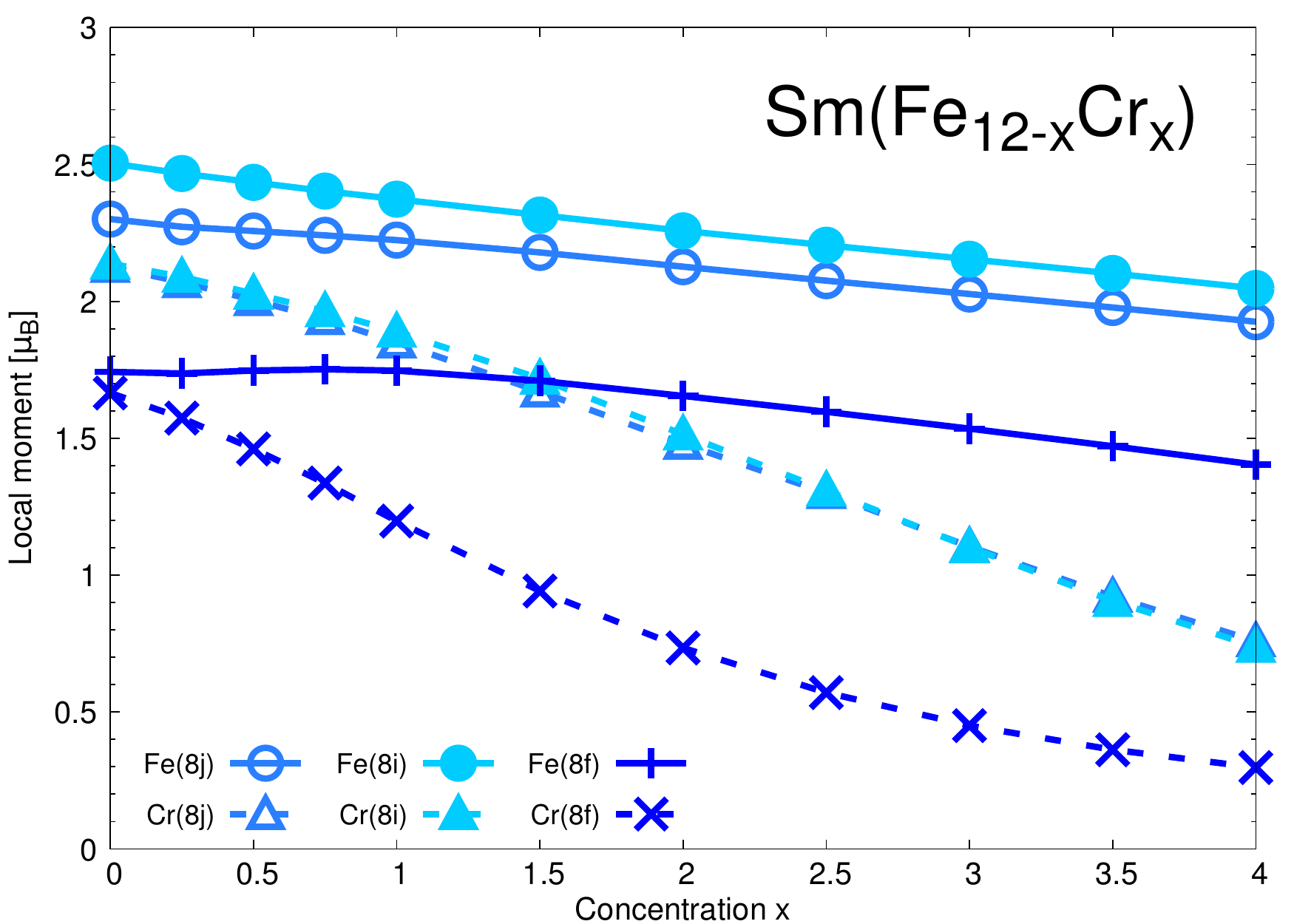}
  \vspace{1cm}
  \caption{
  (Color online)
  Absolute values of the local moment at the Fe and Cr sites
  in SmFe$_{12-x}$Cr$_x$. All the Cr moments are antiparallel to 
  the direction of the Fe moments.
  \label{local_moments_Sm}
  }
 \end{figure}

The weakening of the Fe--Cr couplings can be related to reduction of the local moment at the Cr sites. 
As has been discussed previously\cite{Akai98,Akai06},
the antiferromagnetic coupling of Cr to the surrounding Fe elements is
energetically stable due to hybridization between states in the $d$-bands
energetically close to each other.
On the other hand,
Cr prefers to couple itself antiparallel with 
the surrounding Cr elements due to hybridization between
energetically separated states (or superexchange)\cite{Akai98,Akai06}.
However, this is against the Fe-Cr coupling that favors Cr pairs to couple ferromagnetically.
Instead of being totally against it,
Cr 
reduces its local moment (and 
sacrifices the intra-atomic exchange energy)
to relax the band energy with hybridization
when the concentration of Cr increases.
Therefore, increase of the Cr concentration results in the reduction of the
Cr moment shown in Fig. \ref{local_moments_Sm}.
The Fe--Cr coupling simultaneously becomes weaker 
as shown in Fig. \ref{J_ave}.

\section{Conclusion}
We calculated the electronic structure of RFe$_{12-x}$Cr$_x$ and 
RFe$_{12-x}$Co$_x$ based on first-principles.
Intersite magnetic couplings and the Curie temperature 
were also calculated using the mean-field approximation.
Our results predict the enhancement in Curie temperature 
through the introduction of Cr in RFe$_{12}$.
The optimal concentration appears to fall between 
$x=0.5$ and $x=0.75$, and the gain is approximately 100~K.

Moreover, Cr is found to be more efficient than Co in enhancing the Curie temperature 
of RFe$_{12}$ in the $x\ll 1$ regime.
We analyzed this feature by decomposing $dT_\text{C}/dx$, the gradient of 
the Curie temperature with respect to the concentration at $x=0$, into two parts.
The Direct part represents the effect of replacing the Fe--Fe couplings with Fe--Cr couplings;
the Indirect part represents the influence of Cr on the remaining Fe--Fe couplings. 
This decomposition relates to an idea previously discussed by Ogura et al.\cite{Ogura11}, specifically that Cr can enhance the magnetism of the Fe--Fe sub-lattices by making nearby Fe atoms appear Co-like because it is attributed only to the Indirect part if this effect can enhance the Curie temperature.

In our calculation, the contribution from the Direct part was found to be almost as equally important as the Indirect part, which means the Fe--Cr couplings play important roles in the enhancement of $T_\text{C}$.
Moreover, these Fe--Cr couplings were found to weaken as the concentration of Cr increases.
We suggest that this weakening and the antiferromagnetic nature of Cr--Cr couplings may explain why Cr can enhance the Curie temperature only when the concentration is small. 

\nocite{suppl}

\begin{acknowledgments}
The authors gratefully acknowledge the support from the Elements Strategy Initiative Project under the auspices of MEXT. This work was also supported by MEXT as a social and scientific priority issue (Creation of new functional Devices and high-performance Materials to Support next-generation Industries; CDMSI) to be tackled by using the post-K computer. 
The computation was partly conducted using the facilities of the Supercomputer Center, the Institute for Solid State Physics, the University of Tokyo, and the supercomputer of ACCMS, Kyoto University. 
This research also used computational resources of the K computer provided by the RIKEN Advanced Institute for Computational Science through the HPCI System Research project (Project ID:hp170100). 
\end{acknowledgments}

\appendix
\section{Lattice parameters}
\label{lattparams}
Table \ref{table2} lists the lattice parameter settings we used in the calculations.
As described in the above, we use the lattice parameters of
(a) YFe$_{12}$ for YFe$_{12-x}$Cr$_x$ and YFe$_{12-x}$Co$_x$,
(b) NdFe$_{12}$ for NdFe$_{12-x}$Cr$_x$ and NdFe$_{12-x}$Co$_x$, and 
(c) SmFe$_{12}$ for SmFe$_{12-x}$Cr$_x$ and SmFe$_{12-x}$Co$_x$.
We assumed the ThMn$_{12}$ structure [space group: I4/mmm (\#139)]
for the systems.
The definitions of $p_\text{8i}$ and $p_\text{8j}$ are summarized in 
Table \ref{table1} with representable atomic positions of the atoms.
\begin{table}[h]
  \caption{
 Optimized lattice parameters for RFe$_{12}$ (R = Y, Nd, Sm).
 See Table \ref{table1} for definitions of $p_{\rm 8i}$ and $p_{\rm 8j}$.
  \label{table2}}
 \begin{tabular}{ccccc}
  \hline
  \hline
   R & $a$ [\AA] & $c$ [\AA] &$p_{\rm 8i}$& $p_{\rm 8j}$\\
  \hline
                          Y &  8.453 &   4.691 & 0.3583 & 0.2721 \\
                         Nd &  8.533 &   4.681 & 0.3594 & 0.2676 \\
                         Sm &  8.497 &   4.687 & 0.3588 & 0.2696 \\
  \hline
  \hline
 \end{tabular}
 \end{table}
\begin{table}[h]
 \caption{Atomic positions for RFe$_{12}$ (R = Y, Nd, Sm). The variables, $x$, $y$, and $z$, denote the point ($ax$, $ay$, $cz$) in Cartesian coordinates.
 \label{table1}
 }
 \begin{tabular}{ccrrr}
  \hline
  \hline
  Element & Site & $x$ & $y$& $z$\\
  \hline
  Nd & 2a &            0 &    0 &    0 \\
  Fe & 8f &         0.25 & 0.25 & 0.25 \\
  Fe & 8i & $p_{\rm 8i}$ &    0 &    0 \\
  Fe & 8j & $p_{\rm 8j}$ &  0.5 &    0 \\
  \hline
  \hline
 \end{tabular}
 \end{table}

\section{Magnetization}
\begin{figure}[htbp]
  \center
  \includegraphics[width=8cm]{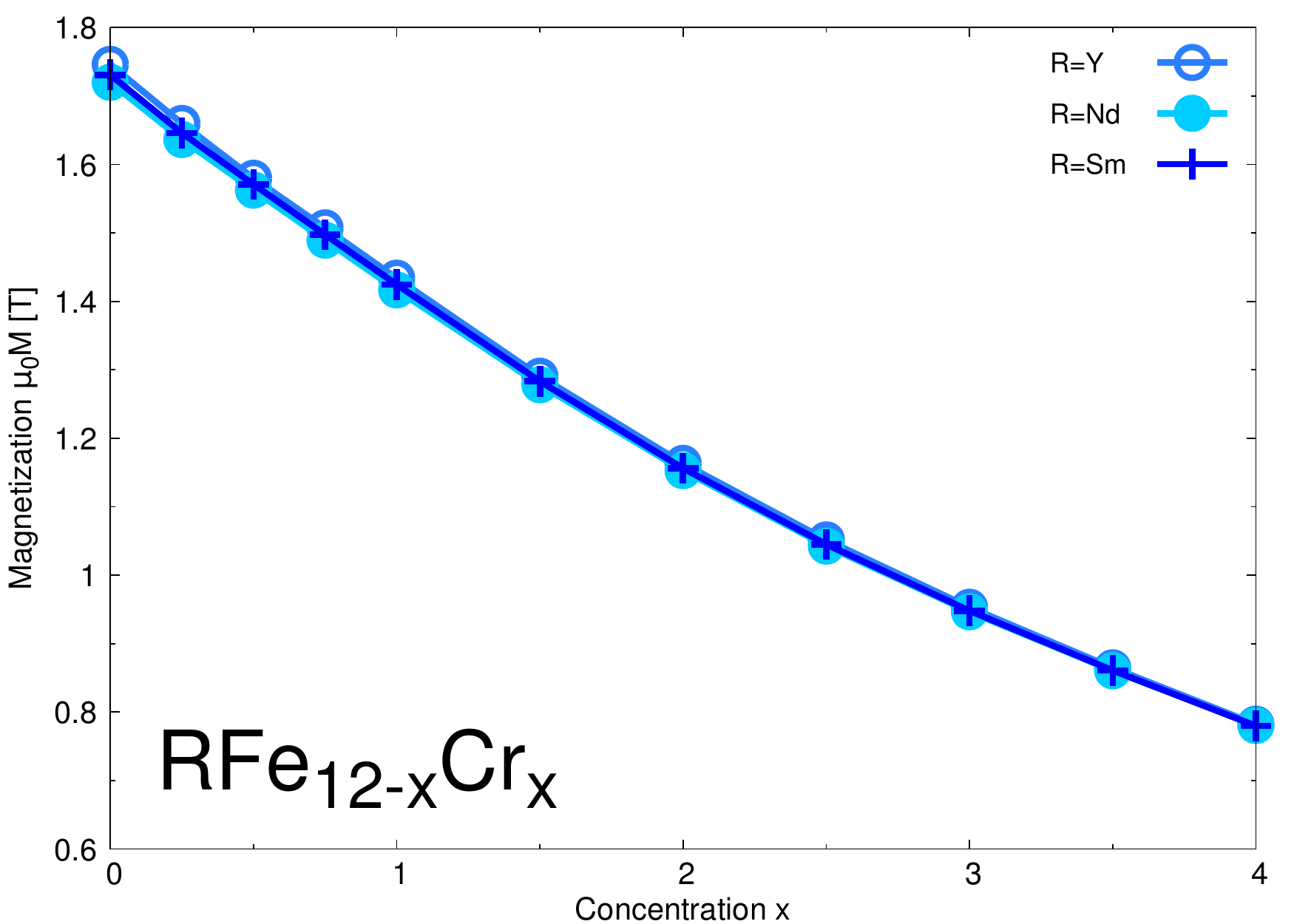}
  \vspace{1cm}
  \caption{
  (Color online)
  Magnetization values for RFe$_{12-x}$Cr$_{x}$ as functions of Cr concentration $x$.
  The contribution from the R-f electrons are excluded from those values.
  \label{Magnetization}
  }
 \end{figure}

Figure \ref{Magnetization} plots the calculated values of magnetization for RFe$_{12-x}$Cr$_{x}$ as functions of Cr concentration $x$ in the range of $0\leq x \leq 4$.
The contribution from the R-f electrons are excluded from those values.
The magnetization is significantly reduced with the introduction of Cr mainly because Cr has an antiparallel magnetic moment to the Fe moments.

\section{Conversion of the intersite magnetic coupling to a Curie temperature}
\label{section_MFA}
We here summarize how we apply the mean-field approximation to the Hamiltonian given as Eq.~\eqref{HeisenbergHamiltonian} and \eqref{RandomizedJij} to obtain the Curie temperature.
The methodology is essentially identical with that used by previous authors for their problems (e.g., [\citen{Sato59,Oguchi69,Matsubara77}]).

We consider the fluctuation of $\vec{e}_{i,n}$ from the sample average assuming it to be sufficiently small.
However, the nature of $\vec{e}_{i,n}$ strongly depends on the atom that occupies the site (e.g., Fe or Cr as in the main text).
To avoid this problem, we introduce an extra spin $\vec{e}^\text{\,A}_{i,n}$ associated with atom A and make the replacement
\begin{equation}
 \vec{e}_{i,n} \rightarrow
  \sum_\text{A} \gamma^\text{A}_i(n) \vec{e}^\text{\,A}_{i,n}.
\end{equation}
Because $\gamma^\text{A}_i(n) = 1$ only when A is the atom that occupies the $i$th site in the $n$th sample and vanishes otherwise, this does not change the physical meaning of the Hamiltonian of Eq.~\eqref{HeisenbergHamiltonian}.
With this substitution, one can rewrite the Hamiltonian as
\begin{equation}
 \mathcal{H}(n)
  =
  -
  \sum_{i,j}
  \sum_{\rm A, B}
  J^{\text {A--B}}_{i, j}\,
  \left\{
  \gamma^{\rm A}_i(n)
  \,
  \vec{e}^\text{\,A}_{i,n}
  \right\}
  \cdot
  \left\{
  \gamma^{\rm B}_j(n)
  \,
  \vec{e}^\text{\,B}_{j,n}
  \right\}
  .
  \label{HeisenbergHamiltonian2}
\end{equation}

We now consider deviations of $\vec{e}^\text{\,A}_{i,n}$ from the double (thermal and sample) average of itself.
Let $\vec{\chi}^\text{A}_{i} \equiv\llangle \vec{e}^\text{\,A}_{i,n} \rrangle$ denote this average.
We also consider the fluctuation of the $\gamma^{\rm A}_i(n)$ from its sample average,
$C^{\rm A}_i$, which we may call the concentration.
These fluctuations, $\delta \vec{e}^\text{\,A}_{i,n}$ and 
$\delta\gamma^{\rm A}_i(n)$, are defined as follows:
\begin{align}
 &\vec{e}^\text{\,A}_{i,n}
  = \vec{\chi}^\text{A}_{i}
 + \delta \vec{e}^\text{\,A}_{i,n},\\
 &\gamma^{\rm A}_i(n) = C^{\rm A}_i + \delta\gamma^{\rm A}_i(n).
\end{align}
We need to treat the correlation between $\delta e$ and $\delta \gamma$ at a site.
By noticing 
$\llangle \vec{e}^\text{\,A}_{i,n} \rrangle =
\llangle  \gamma^\text{\,A}_i(n) \vec{e}^\text{\,A}_{i,n}\rrangle$,
one can show
$\llangle \delta\gamma^\text{\,A}_i(n)
\delta\vec{e}^\text{\,A}_{i,n} \rrangle
= (1-C^\text{A}_i)\vec{\chi}^\text{A}_i$.
Therefore, $\gamma^\text{\,A}_i(n) \vec{e}^\text{\,A}_{i,n}$
in Eq.~\eqref{HeisenbergHamiltonian2} can be expressed as
\begin{align}
 \gamma^\text{\,A}_i(n) \vec{e}^\text{\,A}_{i,n}
  &=
  \vec{\chi}^\text{A}_i
  +
  \delta \gamma^\text{\,A}_i(n) \vec{\chi}^\text{A}_i
  +
  \delta\vec{e}^\text{\,A}_{i,n}    C^\text{A}_i
  \nonumber \\
  &\quad
  +
  \delta\gamma^\text{\,A}_i(n) \delta\vec{e}^\text{\,A}_{i,n}
  -
  \llangle
      \delta\gamma^\text{\,A}_i(n)
      \delta\vec{e}^\text{\,A}_{i,n}
  \rrangle
 \\
 &\equiv
  \vec{\chi}^\text{A}_i
  +
 \delta
 \left[
 \gamma^\text{\,A}_i(n) \vec{e}^\text{\,A}_{i,n}
 \right],
\end{align}
wherein the defined $\delta\left(\gamma\, \vec{e}\,\right)$ satisfies
$ \llangle
 \delta
 \left[
 \gamma^\text{\,A}_i(n) \vec{e}^\text{\,A}_{i,n}
 \right] \rrangle = 0$.
 
By omitting the constant term and the terms with
$\delta\left(\gamma\, \vec{e}\,\right)^2$,
one can obtain the approximate Hamiltonian,
\begin{equation}
  H(n)
  =
  -2
  \sum_{i, \text{A}}
  \left\{
  \gamma^{\rm A}_i(n)
  \,
  \vec{e}^\text{\,A}_{i,n}
  \right\}
  \cdot
  \sum_{j, \text{B}}
  J^{\text {A--B}}_{i, j}\,
  \vec{\chi}^\text{\,B}_j
  .
  \label{HeisenbergHamiltonian3}
\end{equation}
As this is simply the mean-field Hamiltonian of the ordinary Heisenberg model,
a self-consistent equation can be obtained,
\begin{equation}
 \vec{h}^\text{A}_i
  =
  \sum_{j,\text{B}}
  2
  J^{\text {A--B}}_{i, j}
  C^\text{B}_j
  L\left(
    \beta\left|\vec{h}^\text{B}_j\right|
    \right)
  \frac{\vec{h}^\text{B}_j}
       {\left|\vec{h}^\text{B}_j\right|},
  \label{self-consistent_eq}
\end{equation}
where $\vec{h}^\text{A}_i$ is related to $\vec{\chi}$ by
\begin{equation}
\vec{\chi}^\text{\,A}_i
  =
  L\left(
    \beta\left|\vec{h}^\text{A}_i\right|
    \right)
  \frac{\vec{h}^\text{A}_i}
       {\left|\vec{h}^\text{A}_i\right|},
\end{equation}
$\beta=1/k_\text{B}T$---the inverse of temperature divided by the Boltzmann constant---and $L(x)$ is the Langevin function.
The Curie temperature is the supremum of the $\beta$ values with which
Eq.~\eqref{self-consistent_eq} can have a nontrivial solution.

In solving Eq.~\eqref{self-consistent_eq} for the temperature, we use
$L\left(\beta\left|\vec{h}^\text{A}_i\right|\right)\sim
\beta\left|\vec{h}^\text{A}_i\right|/3$,
an asymptotic function of $L(x)$ in the limit of $x\rightarrow 0$,
which is accurate when $\left|\vec{\chi}^\text{A}_i\right| =
L\left(\beta\left|\vec{h}^\text{A}_i\right|\right)$ is small.
The resulting equation is 
\begin{equation}
  \sum_{j,\text{B}}
 \left(
  \delta_{i,j}\delta_\text{A,B}
  -
  \frac{2\beta}{3}
  J^{\text {A--B}}_{i, j}
  C^\text{B}_j
 \right)
  \vec{h}^\text{B}_j
  =
  \vec{0}.
\end{equation}
This equation has a nontrivial solution at $T=2\lambda/3k_\text{B}(\equiv T_\text{C})$ where $\lambda$ is the largest eigenvalue of $P=\left[J^{\text {A--B}}_{i, j} C^\text{B}_j\right]$, the matrix that has $J^{\text {A--B}}_{i, j} C^\text{B}_j$ as an element.

It can also be proved that there is no solution other than the trivial one to Eq.~\eqref{self-consistent_eq} for $T\geq T_\text{C}$ (or $2\lambda\beta/3\leq 1$) as follows.
The Langevin function satisfies the inequality $L\left(\beta\left|\vec{h}^\text{A}_i\right|\right) \leq\beta\left|\vec{h}^\text{A}_i\right|/3$, where equality holds only when $\left|\vec{h}^\text{A}_i\right|=0$. The matrix
$Q=
  \left[
  2J^{\text {A--B}}_{i, j}
  C^\text{B}_j\,
  L\left(
           \beta\left|\vec{h}^\text{B}_j\right|
           \right)
       /\left|\vec{h}^\text{B}_j\right|
 \right]
$, the elements of which appear in Eq.~\eqref{self-consistent_eq}, is related to $P$ by $(2\beta/3)P=QD^2$ where $D=\left[\delta_{i,j}\,\delta_\text{A,B}
    \sqrt{
       \beta\left|\vec{h}^\text{B}_j\right|
       /
       3L\left(
           \beta\left|\vec{h}^\text{B}_j\right|
           \right)
    }
   \right]
$. This matrix $D$ satisfies $\left|D\vec{u}\right| \geq \left|\vec{u}\right|$ for any $\vec{u}$ because $D$ is diagonal and all the diagonal elements are larger than or equal to 1.
Therefore, the largest eigenvalue $\kappa$ of the matrix Q is
\begin{align}
 \kappa &\leq 
 \max_{\vec{v}}
 \frac{1}{\left|\vec{v}\right|^2}
 \vec{v}^\dagger
 Q
 \vec{v}
 =
 \max_{D\vec{u}}
 \frac{1}{\left|D\vec{u}\right|^2}
 \vec{u}^\dagger
 D
 Q
 D
 \vec{u} \nonumber \\
 &\leq
 \max_{\vec{u}}
 \frac{1}{\left|\vec{u}\right|^2}
 \vec{u}^\dagger
 D
 Q
 D
 \vec{u}
 =\frac{2\beta\lambda}{3},
\end{align}
where the last equality comes from $\det(\gamma I-QD^2)=\det(D)\det(\gamma I-QD^2)\det(D^{-1})=\det(\gamma I - DQD)$, which holds for any $\gamma$.
Therefore, $I-Q$ is positive definite, which means there is no non-trivial solution to Eq.~\eqref{self-consistent_eq}.

\bibliographystyle{jpsj.bst}
\bibliography{nasu}

\begin{thebibliography}{10}

\bibitem{Wang93}
D.~Wang, S.-H. Liou, P.~He, D.~J. Sellmyer, G.~Hadjipanayis, and Y.~Zhang:
  Journal of magnetism and magnetic materials {\bfseries 124} (1993) 62.

\bibitem{Hirayama15}
Y.~Hirayama, Y.~Takahashi, S.~Hirosawa, and K.~Hono: Scripta Materialia
  {\bfseries 95} (2015) 70.

\bibitem{Hirayama15b}
Y.~Hirayama, T.~Miyake, and K.~Hono: JOM {\bfseries 67} (2015) 1344.

\bibitem{Buschow91}
K.~Buschow: Journal of magnetism and magnetic materials {\bfseries 100} (1991)
  79.

\bibitem{Ogura11}
M.~Ogura, H.~Akai, and J.~Kanamori: Journal of the Physical Society of Japan
  {\bfseries 80} (2011) 104711.

\bibitem{Hao96}
Y.~M. Hao, P.~L. Zhang, J.~X. Zhang, X.~D. Sun, Q.~W. Yan, Ridwan, Mujamilah,
  Gunawan, and Marsongkohadi: Journal of Physics: Condensed Matter {\bfseries
  8} (1996) 1321.

\bibitem{Girt97}
E.~Girt, Z.~Altounian, and J.~Yang: Journal of applied physics {\bfseries 81}
  (1997) 5118.

\bibitem{Liechtenstein87}
A.~I. Liechtenstein, M.~Katsnelson, V.~Antropov, and V.~Gubanov: Journal of
  Magnetism and Magnetic Materials {\bfseries 67} (1987) 65.

\bibitem{Kohn65}
W.~Kohn and L.~J. Sham: Physical Review {\bfseries 140} (1965) A1133.

\bibitem{Hohenberg64}
P.~Hohenberg and W.~Kohn: Physical Review {\bfseries 136} (1964) B864.

\bibitem{Perdew81}
J.~P. Perdew and A.~Zunger: Phys. Rev. B {\bfseries 23} (1981) 5048.

\bibitem{Fukazawa17}
T.~Fukazawa, H.~Akai, Y.~Harashima, and T.~Miyake: Journal of Applied Physics
  {\bfseries 122} (2017) 053901.

\bibitem{QMAS}
{QMAS|Quantum MAterials Simulator Official Site}.
\newblock http://qmas.jp.

\bibitem{Bloechl94}
P.~E. Bl\"ochl: Phys. Rev. B {\bfseries 50} (1994) 17953.

\bibitem{Kresse99}
G.~Kresse and D.~Joubert: Phys. Rev. B {\bfseries 59} (1999) 1758.

\bibitem{Harashima14b}
Y.~Harashima, K.~Terakura, H.~Kino, S.~Ishibashi, and T.~Miyake: Proceedings of
  Computational Science Workshop 2014 (CSW2014), Vol.~5 of {\em JPS Conference
  Proceedings}, 2015, p. 1021.

\bibitem{Hirayama17}
Y.~Hirayama, Y.~Takahashi, S.~Hirosawa, and K.~Hono: Scripta Materialia
  {\bfseries 138} (2017) 62.

\bibitem{Akai98}
H.~Akai: Physical Review Letters {\bfseries 81} (1998) 3002.

\bibitem{Akai06}
H.~Akai and M.~Ogura: Physical review letters {\bfseries 97} (2006) 026401.

\bibitem{suppl}
(Supplemental material) detailed data from our calculation: the total and local
  magnetic moments, the average intersite magnetic interaction with wider range
  of the Cr concentration, and partial d-DOS of the Cr and Fe sites in
  Sm(FeCr)$_{12}$ are provided online.

\bibitem{Sato59}
H.~Sato, A.~Arrott, and R.~Kikuchi: Journal of Physics and Chemistry of solids
  {\bfseries 10} (1959) 19.

\bibitem{Oguchi69}
T.~Oguchi and T.~Obokata: Journal of the Physical Society of Japan {\bfseries
  27} (1969) 1111.

\bibitem{Matsubara77}
F.~Matsubara and S.~Inawashiro: Journal of the Physical Society of Japan
  {\bfseries 42} (1977) 1529.

\end{thebibliography}
\end{document}